\documentclass[a4paper,11pt]{article}
\usepackage{pos}
\usepackage{wrapfig}

\title{Searches for TeV emission from Northern Hemisphere pulsars with VERITAS}

\author*[1]{Samantha L. Wong}
\author[]{for the VERITAS collaboration} 

\author[1]{Dinah Ibrahim}
\author[2]{Benjamin Shaw}

\affiliation[1]{Physics Department, McGill University, Montreal, QC H3A 2T8, Canada}
\affiliation[2]{Jodrell Bank Centre for Astrophysics, School of Physics and Astronomy, The University of Manchester, Manchester M13 9PL, UK}

\emailAdd{samantha.wong2@mail.mcgill.ca}

\abstract{The discovery of VHE emission from the Crab pulsar and, more recently, multi-TeV emission from the Vela pulsar have challenged our current understanding of the emission mechanisms of these sources. Studying pulsar emission at TeV energies allows us to understand the engines that power some of the most extreme accelerators in the Galaxy. We present recent highlights from the VERITAS pulsar program using nearly two decades of VERITAS data and novel high energy analysis techniques optimized for emission up to 100 TeV. This work begins to characterize how the emerging population of multi-TeV pulsars can be predicted from existing multi-wavelength observations. In particular, we highlight a search for VHE emission above 1 TeV using over 17 years of Crab pulsar data, which extends the high energy end of the existing VERITAS spectrum. Additionally, we search for both optical and multi-TeV emission from bright Vela-like pulsars, including analysis of over 200 hours of data on PSR J2229+6114, which powers the Boomerang pulsar wind nebula and is putatively associated with the ultra-high-energy source 1LHAASO J2229+5927u. We discuss these results in the context of the broader pulsar population and their impacts on the prospects of new pulsar discoveries with next-generation VHE instruments.}

\ConferenceLogo{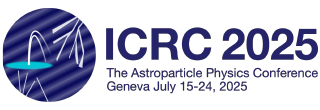}

\FullConference{39th International Cosmic Ray Conference (ICRC2025)\\
 15–24 July 2025\\
Geneva, Switzerland\\}

\begin{document}

\maketitle

\section{Introduction}
Though there are over 3000 pulsars detected across the electromagnetic spectrum \cite{atnf} and 294 detected by the \textit{Fermi}-LAT \cite{3pc}, only the Crab \cite{vtscrab} and Vela \cite{hessvela} pulsars have been confirmed to emit pulsed emission at $>5\sigma$ statistical significance above 1 TeV. The Crab and Vela pulsars exhibit distinct spectral shapes, with the Crab pulsar interpulse (P2) described by a power law with a soft spectral index ($\Gamma = 3.0 \pm 0.1$), which falls below IACT sensitivity just above 1 TeV \cite{magiccrab}. In contrast, the Vela pulsar follows a very hard power law spectrum ($\Gamma = 1.4 \pm 0.3$) \cite{hessvela}. Both the Crab and Vela pulsar are extraordinarily energetic and bright sources, but the question remains as to where to look for the next TeV pulsar candidate.

In this work, VERITAS reports a non-detection of MAGIC's 1 TeV Crab pulsar (P2) spectral point \cite{magiccrab} with a high quality, 18-year dataset and a non-detection from the deep analysis of a bright, Vela-like pulsar PSR J2229+6114, which was chosen as a candidate to search for the next TeV-emitting pulsar. 

VERITAS is an array of four imaging atmospheric Cherenkov telescopes (IACTs) located at the Fred Lawrence Whipple Observatory (FLWO) in southern Arizona (31 40N, 110 57W,  1.3km a.s.l.). VERITAS has a rich history of pulsar studies, from the initial detection of Crab pulsar emission above 100 GeV \cite{vtscrab} to an archival search for emission from thirteen young pulsars \cite{archivalpsr}. These efforts, along with continued observations of pulsar-powered Galactic sources, such as pulsar wind nebulae (PWNe) and TeV halos, have allowed VERITAS to accumulate a total of 25 000 hours of data on Northern Hemisphere pulsars\footnote{Note that there are many cases where several pulsars fall within the same field of view, so some exposures are double (or more) counted.}.

\section{VERITAS analysis of the Crab Pulsar}
Following MAGIC's detection of $> 1$ TeV emission from the Crab pulsar's interpulse \cite{magiccrab}, efforts have been made by VERITAS to confirm MAGIC's data points (e.g., \cite{nguyen_crab}). The detection of pulsed emission up to 1.5 TeV necessitates a Lorentz factor greater than $5 \times 10^6$ \cite{magiccrab}, which can only originate from leptonic emission near or beyond the light cylinder and strongly constrains  magnetospheric models \cite{bogovalov} that are traditionally used to explain pulsed emission up to GeV energies (e.g., \cite{polarcap}). 

In this work, VERITAS has performed a new Crab pulsar analysis using 329 hours of data, taken between 2007 and 2025. Because the Crab pulsar is extremely background dominated, both by hadronic background events and by Crab nebula emission, it is critical to maximize the pulsar's signal-to-noise ratio. To this end, strong quality cuts are imposed to remove any data taken with $<4$ telescopes, large zenith angle ($> 40^\circ$), or taken with a mean energy threshold above 400 GeV. These quality cuts improve the signal-to-noise ratio for the Crab pulsar interpulse by $>$10\%. Event reconstruction and gamma/hadron separation for this analysis were performed with standard VERITAS tools \cite{edref,vegref}, while the spectrum was obtained using \textit{Gammapy}'s phase background maker \cite{gammapy:2023,acero_2025_14760974}, which allows for temporal separation of events to avoid spectral contamination by the Crab nebula (which has not been found to vary temporally in the VHE regime; e.g., \cite{vanscherpenberg2019searchingvariabilitycrabnebula}). 

\begin{figure}[h]
    \centering
    \includegraphics[width=0.7\linewidth]{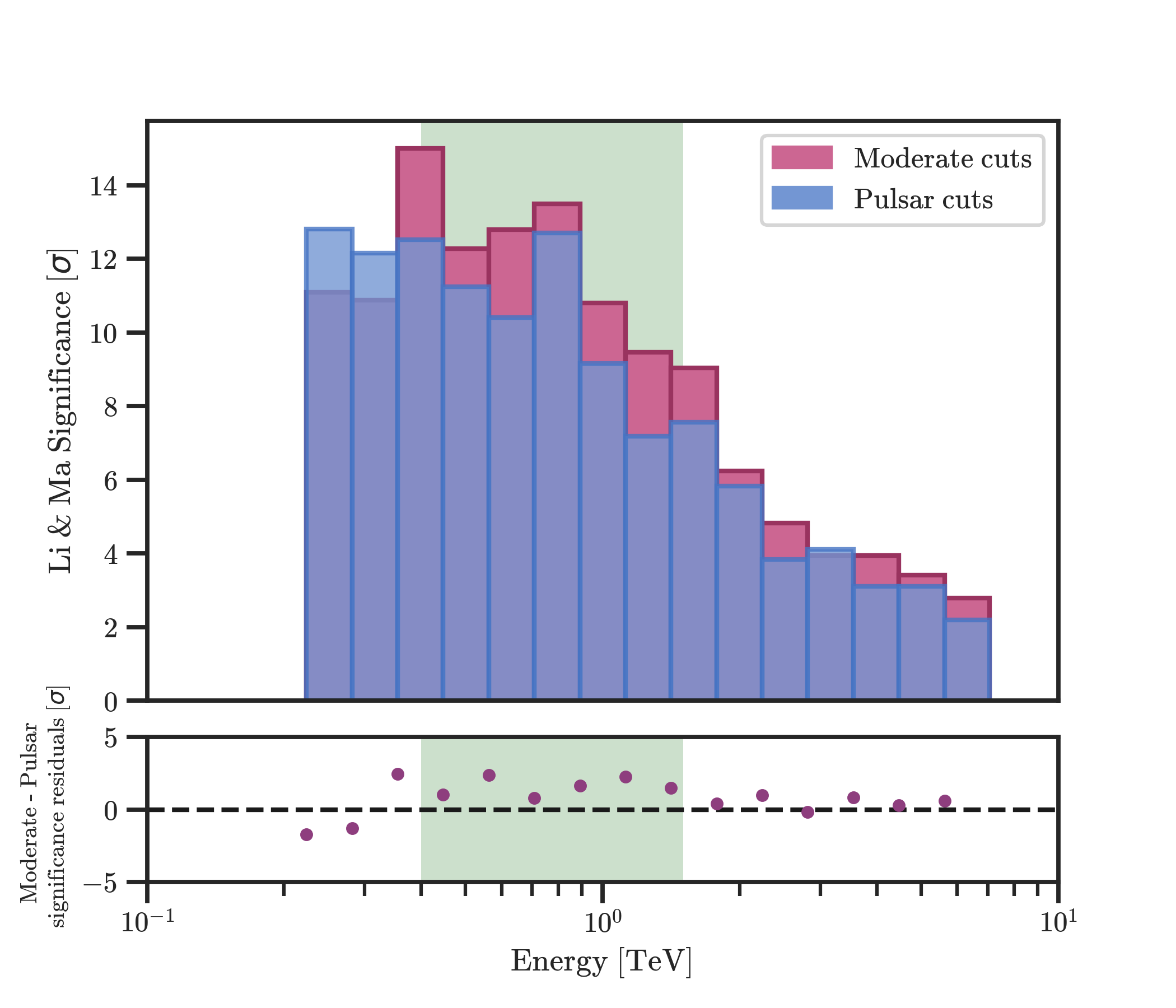}
    \caption{Comparison of spectral bin significances (calculated with Eq.17 from \cite{lima}) for Crab nebula data, using both moderate gamma/hadron cuts and pulsar gamma/hadron cuts. The energy range of interest for this work is shaded in green.}
    \label{fig:cutcompare}
\end{figure}

All gamma-ray-like events are assigned temporal ON/OFF regions (within the spatial ON region) based on the phase gates defined in the original Crab pulsar detection \cite{vtscrab}. These events are barycentered and assigned phases using the pulsar timing package TEMPO2 \cite{2006MNRAS.369..655H} and the Jodrell Bank Crab pulsar monthly ephemeris\footnote{http://www.jb.man.ac.uk/~pulsar/crab.html} \cite{1993MNRAS.265.1003L}. 

This analysis revisits the gamma/hadron separation cuts used in previous analyses \cite{vtscrab,nguyen_crab}, which used ``pulsar cuts'' to remove signal contamination from the Crab nebula by optimizing Monte Carlo simulated events to produce a very soft power law spectrum ($\Gamma = 4$). In the energy range of interest for this work (400 GeV - 1.5 TeV), standard VERITAS moderate cuts - optimized for the Crab nebula spectrum ($\Gamma \approx 2.4$) - are more sensitive (see Fig. \ref{fig:cutcompare}), given that the Crab nebula emission contribution is well-estimated by the phase-based background calculation. Further cut optimization does not benefit the pulsar detection at 1 TeV, since the images of 1 TeV gamma-rays from both the Crab nebula and pulsar should be indistinguishable. 

\begin{figure}
    \centering
    \includegraphics[width=0.7\linewidth]{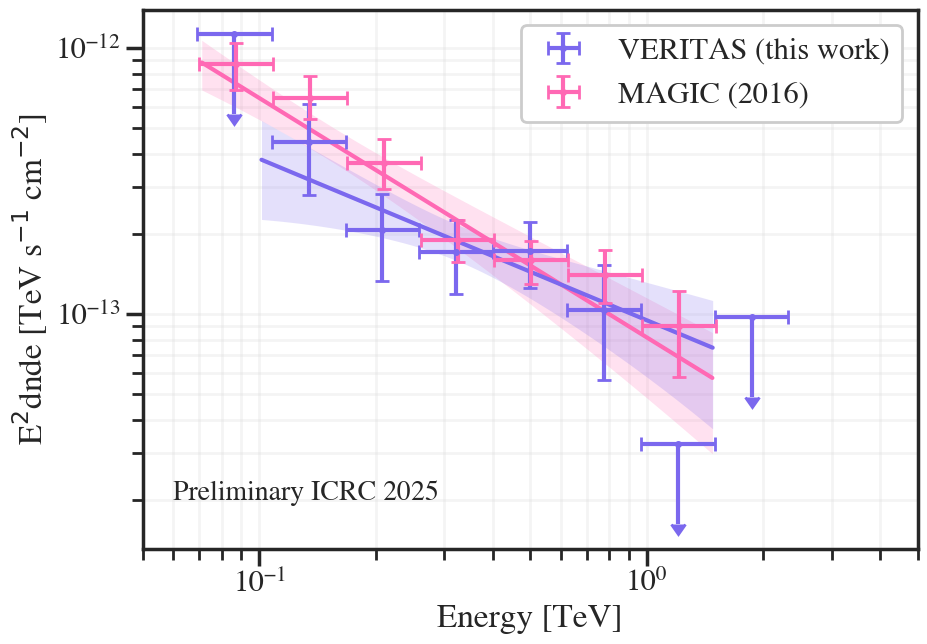}
    \caption{VERITAS Crab pulsar (P2) spectrum compared with MAGIC's spectrum \cite{magiccrab}. Upper limits are 95\% C.L., calculated with forward-folded Li \& Ma \cite{lima} significance, assuming the MAGIC spectral model.}
    \label{fig:crabspec}
\end{figure}

Though the overall significance of pulsed emission is expected to decrease compared to the original "pulsar cuts" (since the signal is dominated by $< 200$ GeV emission, below the energy threshold of moderate cuts), the Crab pulsar interpulse (P2) is re-detected at 5$\sigma$ in this analysis. The spectral energy distribution obtained in this analysis provides an additional spectral point centered at 781 GeV (see Fig. \ref{fig:crabspec}), building on the 194 hours of data analyzed in previous work \cite{nguyen_crab}. Although this spectral point is consistent with similar points from both MAGIC and LST-1 \cite{magiccrab,lstcrab}, VERITAS remains unable to reconstruct a statistically significant spectral point above 1 TeV (see Table \ref{ref:fluxtab}). The VERITAS spectrum is best-described by a power law with a spectral index $\Gamma = 2.6 \pm 0.32$ and normalization $\Phi_0 = (1.33 \pm 0.39) \times10^{-11} \textrm{cm}^2 \textrm{ s}^{-1} \textrm{ TeV}^{-1}$ at a reference energy of 0.15 TeV, which was chosen to match MAGIC's reference energy in \cite{magiccrab}. We optimized the selection of runs used in this analysis by selecting only those with effective areas that would allow for the detection of the 1 TeV flux point within the VERITAS exposure. These effective areas are at least equal to those shown for the high zenith angle sample in MAGIC's analysis (Figure 1 of \cite{magiccrab}). Further work is underway to investigate whether the significance can be improved by modifying equation 17 of \cite{lima} to incorporate spatial OFF regions (in addition to the existing temporal OFF regions) to improve the estimate of the Crab nebula excess, which can be subtracted to improve the signal-to-noise ratio in higher energy, low count bins. 

\begin{table}[]
\centering
\scalebox{0.6}{
\begin{tabular}{ccccc}
\hline
\hline
\begin{tabular}[c]{@{}c@{}}Energy \\ {[}GeV{]}\end{tabular} & \begin{tabular}[c]{@{}c@{}}Bin Center\\ {[}GeV{]}\end{tabular} & \begin{tabular}[c]{@{}c@{}}P2 E$^2$dnde\\ (MAGIC)\\ {[}10$^{-13}$ TeV$^{-1}$ s$^{-1}$ cm$^{-2}${]}\end{tabular} & \begin{tabular}[c]{@{}c@{}}P2 E$^2$dnde\\ (VERITAS)\\ {[}10$^{-13}$ TeV$^{-1}$ s$^{-1}$ cm$^{-2}${]}\end{tabular} & \begin{tabular}[c]{@{}c@{}}VERITAS significance\\ {[}$\sigma${]}\end{tabular} \\
\hline
69-108 & 87 & 8.7±1.8 &  & -0.59 \\
108-167 & 135 & 6.5±1.2 & 4.4±1.7 & 2.88 \\
167-259 & 210 & 3.7±0.8 & 2.1±0.8 & 2.87 \\
259-402 & 325 & 1.9±0.3 & 1.7±0.5 & 3.33 \\
402-623 & 504 & 1.6±0.3 & 1.7±0.5 & 3.69 \\
623-965 & 781 & 1.4±0.3 & 1.0±0.5 & 2.23 \\
965-1497 & 1211 & 0.9±0.3 & \textless{}0.3 & -1.34 \\
1497-2321 & 1879 & \textless{}0.6 & \textless{}0.9 & -0.13 \\
2321-3598 & 2914 & \textless{}0.8 &  & -2.04\\
\hline
\hline
\end{tabular}
}
\caption{Comparison of VERITAS and MAGIC \cite{magiccrab} spectral points for the measurements shown in Fig.
\ref{fig:crabspec}.}
\label{ref:fluxtab}
\end{table}

The upper limit centered at 1211 GeV appears to be in tension with MAGIC's flux measurement, but it should be noted that this limit is calculated with Li \& Ma significance (Eq. 17 of \cite{lima}), a two-sided distribution which is defined for negative significances, so larger negative excesses will create deeper limits. This is in contrast with Rolke limits \cite{rolke}, which have historically been used in most IACT analyses. The Rolke test statistic is a one-sided distribution that is not defined for negative significances, so negative excess are equivalent to zero excess. Due to the unphysical meaning of a negative excess in this context, we do not claim a tension with MAGIC's spectral point in this energy bin. However, we note that, though VERITAS measures relatively constant significance from 100-965 GeV, the significance in the subsequent bins above 1 TeV falls rapidly to zero (see Table \ref{ref:fluxtab}). 

\section{VERITAS analysis of PSR J2229+6114}
PSR J2229+6114 was selected as the first candidate in a search for Vela-like, TeV emission. PSR J2229+6114 was selected because it is a bright \textit{Fermi}-LAT pulsar (the ninth brightest \textit{Fermi}-LAT pulsar at a declination that can be observed by VERITAS) with a similar multi-wavelength spectral shape to Vela \cite{kuipermwl} and large VERITAS exposure. VERITAS has accumulated 242 hours (after quality cuts) on PSR J2229+6114 through both dedicated observations and observations of SNR G106.6+2.9 (Boomerang), the supernova remnant associated with PSR J2229+6114 \cite{2001ApJ...552L.125H}. 

\begin{figure}[h]
    \centering
    \includegraphics[width=0.7\linewidth]{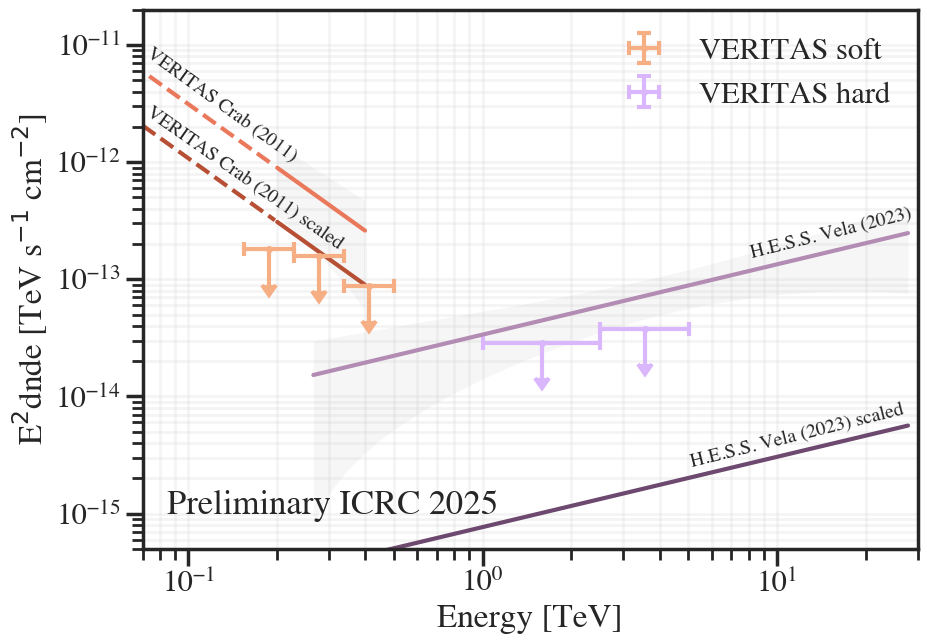}
    \caption{VERITAS flux upper limits for both soft (Crab-like) and hard (Vela-like) emission from PSR J2229+6114. Also shown are the spectral fits of the Crab pulsar by VERITAS \cite{vtscrab} and the Vela pulsar by H.E.S.S. \cite{hessvela}. Additionally, these spectral fits are scaled by the relative \textit{Fermi}-LAT fluxes between the Crab pulsar/Vela pulsar and PSR J2229+6114 to estimate the predicted VHE emission levels.}
    \label{fig:j2229lims}
\end{figure}

In order to properly understand how new TeV pulsars may fit into the Crab-Vela landscape, searches for both hard (Vela-like) and soft (Crab-like) emission were performed on PSR J2229+6114. Both analyses were performed using the analysis and timing packages described above for the Crab pulsar. PSR J2229+6114 ephemerides were provided by the Jodrell Bank Observatory and were calculated for this work in order to account for glitches occurring over the VERITAS observation periods \cite{glitch}\footnote{http://www.jb.man.ac.uk/pulsar/glitches.html}. The results of both analyses can be found in Fig. \ref{fig:j2229lims}. 

First, a search for Vela-like emission was conducted using gamma/hadron cut parameters optimized on a spectral index of 1.4. In order to improve sensitivity to high energy events, this analysis uses an increased ``image loss'' cut \cite{maier_2024_14283615}, which accepts high energy shower images that are typically truncated by VERITAS' relatively small (3.5$^\circ$) field of view. Typically, this image loss cut is implemented in order to remove truncated events, for which the energy and direction are often poorly reconstructed. For this analysis, the threshold for removal of these events is increased from those with 20\% of image pixels at the edge of the camera to removing only those where over 40\% lie at the edge of the camera. This modification results in 1.35$\times$ more excess events for the Crab nebula at 10 TeV and 1.4$\times$ more excess events at 50 TeV. No pulsed emission was detected from this analysis, excluding with 3.6$\sigma$ confidence that PSR J2229+6114 emits at the mean flux level of the Vela pulsar ($\Phi_0 = 1.74 \times 10^{15} \textrm{erg}^{-1} \textrm{cm}^{-2} \textrm{s}^{-1}$ at 4.24 TeV). VERITAS does not have the sensitivity to constrain a Vela-like flux, scaled to the difference between the observed \textit{Fermi}-LAT fluxes in the range 1 GeV $\leq$ E $\leq$ 100 GeV \cite{4fgl}; $F_{\textrm{Vela}}/F_{\textrm{J2229}} = 44$.  

Next, a search for Crab-like emission from PSR J2229+6114 was conducted, using standard cuts optimized for soft spectrum sources. Again, no pulsed emission was detected and emission at both the flux level of the Crab pulsar and the Crab pulsar scaled by the difference in \textit{Fermi}-LAT flux between the Crab pulsar and PSR J2229+6114 (in the range 1 GeV $\leq$ E $\leq$ 100 GeV \cite{4fgl}; $F_{\textrm{Crab}}/F_{\textrm{J2229}} = 2.9$) can be rejected at a level $> 5\sigma$ at 130 GeV. 

\begin{figure}[h]
    \centering
    \includegraphics[width=0.7\linewidth]{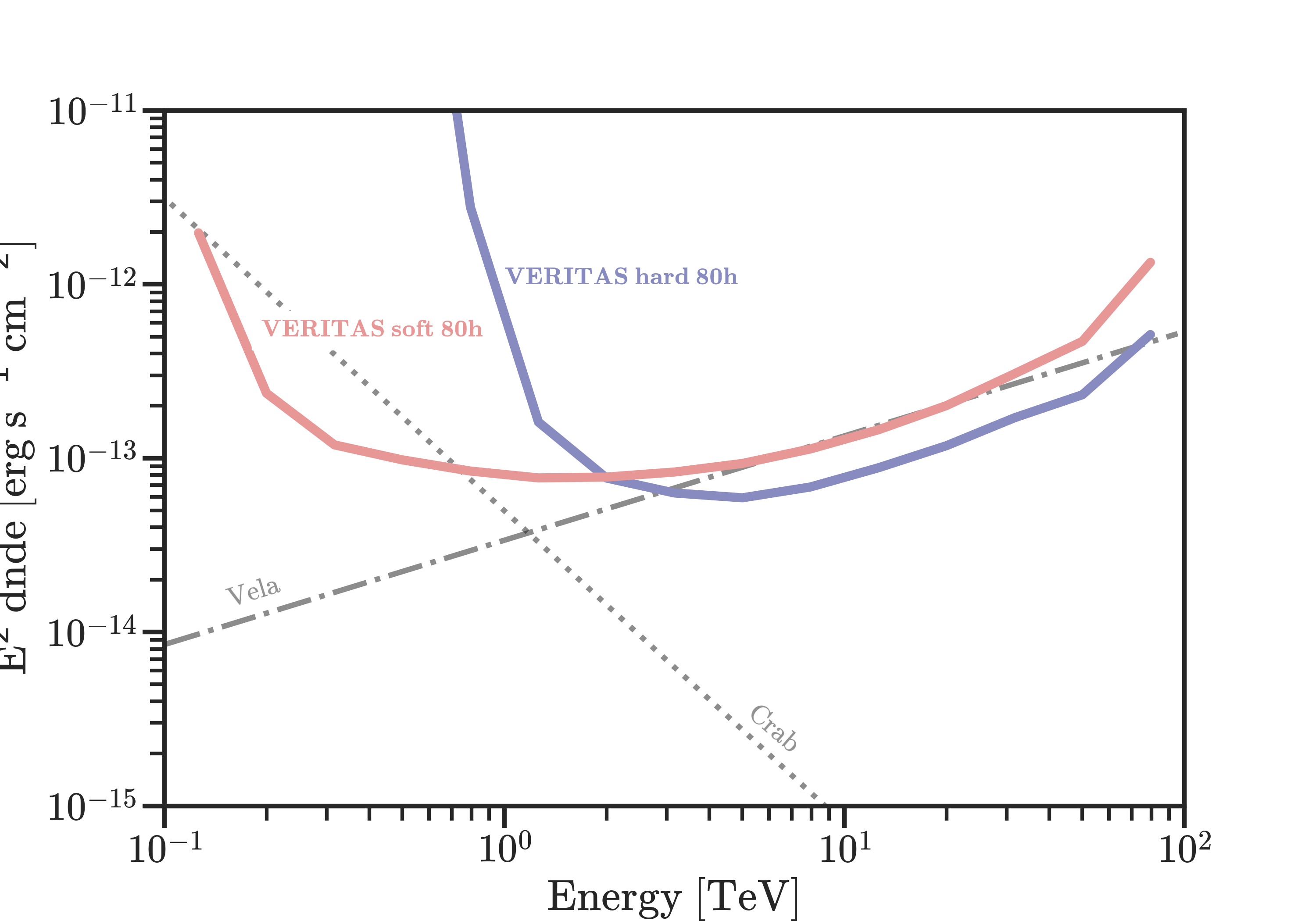}
    \caption{80h VERITAS sensitivity with both soft and hard gamma/hadron cuts. The curve is defined such that each differential energy bin corresponds to a $2 \sigma$ significance, where the integrated significance is 15$\sigma$ between 0.1 TeV $<$ E $<$ 1 TeV for soft cuts and and 8 $\sigma$ between 1 TeV $<$ E $<$ 10 TeV for hard cuts. The Crab pulsar \cite{vtscrab} and Vela pulsar spectra \cite{hessvela} are plotted for reference.}
    \label{fig:sens}
\end{figure}

\section{Future plans: toward a comprehensive Northern Hemisphere pulsar catalog}
Following up the previous VERITAS archival pulsar search \cite{archivalpsr}, this work marks the first steps in producing a comprehensive catalog of all VERITAS pulsars which have at least 80 hours of archival data and would be likely to producing VHE emission, based on multi-wavelength information. This exposure threshold is selected so that both Crab-like and Vela-like spectra can either be detected or strongly constrained (see Fig. \ref{fig:sens}).

The same procedure as described for PSR J2229+6114 in this work will be applied to the selected archival sources, in order to gain insight into which pulsars are likely to be TeV emitters in order to inform searches for pulsed TeV emission by next-generation instruments. 

\section{Acknowledgements}
This research is supported by grants from the U.S. Department of Energy Office of Science, the U.S. National Science Foundation and the Smithsonian Institution, by NSERC in Canada, and by the Helmholtz Association in Germany. This research used resources provided by the Open Science Grid, which is supported by the National Science Foundation and the U.S. Department of Energy's Office of Science, and resources of the National Energy Research Scientific Computing Center (NERSC), a U.S. Department of Energy Office of Science User Facility operated under Contract No. DE-AC02-05CH11231. We acknowledge the excellent work of the technical support staff at the Fred Lawrence Whipple Observatory and at the collaborating institutions in the construction and operation of the instrument.
\clearpage

\section*{Full Author List: VERITAS Collaboration}

\scriptsize
\noindent
A.~Archer$^{1}$,
P.~Bangale$^{2}$,
J.~T.~Bartkoske$^{3}$,
W.~Benbow$^{4}$,
Y.~Chen$^{5}$,
J.~L.~Christiansen$^{6}$,
A.~J.~Chromey$^{4}$,
A.~Duerr$^{3}$,
M.~Errando$^{7}$,
M.~Escobar~Godoy$^{8}$,
J.~Escudero Pedrosa$^{4}$,
Q.~Feng$^{3}$,
S.~Filbert$^{3}$,
L.~Fortson$^{9}$,
A.~Furniss$^{8}$,
W.~Hanlon$^{4}$,
O.~Hervet$^{8}$,
C.~E.~Hinrichs$^{4,10}$,
J.~Holder$^{11}$,
T.~B.~Humensky$^{12,13}$,
M.~Iskakova$^{7}$,
W.~Jin$^{5}$,
M.~N.~Johnson$^{8}$,
E.~Joshi$^{14}$,
M.~Kertzman$^{1}$,
M.~Kherlakian$^{15}$,
D.~Kieda$^{3}$,
T.~K.~Kleiner$^{14}$,
N.~Korzoun$^{11}$,
S.~Kumar$^{12}$,
M.~J.~Lang$^{16}$,
M.~Lundy$^{17}$,
G.~Maier$^{14}$,
C.~E~McGrath$^{18}$,
P.~Moriarty$^{16}$,
R.~Mukherjee$^{19}$,
W.~Ning$^{5}$,
R.~A.~Ong$^{5}$,
A.~Pandey$^{3}$,
M.~Pohl$^{20,14}$,
E.~Pueschel$^{15}$,
J.~Quinn$^{18}$,
P.~L.~Rabinowitz$^{7}$,
K.~Ragan$^{17}$,
P.~T.~Reynolds$^{21}$,
D.~Ribeiro$^{9}$,
E.~Roache$^{4}$,
I.~Sadeh$^{14}$,
L.~Saha$^{4}$,
H.~Salzmann$^{8}$,
M.~Santander$^{22}$,
G.~H.~Sembroski$^{23}$,
B.~Shen$^{12}$,
M.~Splettstoesser$^{8}$,
A.~K.~Talluri$^{9}$,
S.~Tandon$^{19}$,
J.~V.~Tucci$^{24}$,
J.~Valverde$^{25,13}$,
V.~V.~Vassiliev$^{5}$,
D.~A.~Williams$^{8}$,
S.~L.~Wong$^{17}$,
T.~Yoshikoshi$^{26}$\\
\\
\noindent
$^{1}${Department of Physics and Astronomy, DePauw University, Greencastle, IN 46135-0037, USA}

\noindent
$^{2}${Department of Physics, Temple University, Philadelphia, PA 19122, USA}

\noindent
$^{3}${Department of Physics and Astronomy, University of Utah, Salt Lake City, UT 84112, USA}

\noindent
$^{4}${Center for Astrophysics $|$ Harvard \& Smithsonian, Cambridge, MA 02138, USA}

\noindent
$^{5}${Department of Physics and Astronomy, University of California, Los Angeles, CA 90095, USA}

\noindent
$^{6}${Physics Department, California Polytechnic State University, San Luis Obispo, CA 94307, USA}

\noindent
$^{7}${Department of Physics, Washington University, St. Louis, MO 63130, USA}

\noindent
$^{8}${Santa Cruz Institute for Particle Physics and Department of Physics, University of California, Santa Cruz, CA 95064, USA}

\noindent
$^{9}${School of Physics and Astronomy, University of Minnesota, Minneapolis, MN 55455, USA}

\noindent
$^{10}${Department of Physics and Astronomy, Dartmouth College, 6127 Wilder Laboratory, Hanover, NH 03755 USA}

\noindent
$^{11}${Department of Physics and Astronomy and the Bartol Research Institute, University of Delaware, Newark, DE 19716, USA}

\noindent
$^{12}${Department of Physics, University of Maryland, College Park, MD, USA }

\noindent
$^{13}${NASA GSFC, Greenbelt, MD 20771, USA}

\noindent
$^{14}${DESY, Platanenallee 6, 15738 Zeuthen, Germany}

\noindent
$^{15}${Fakult\"at f\"ur Physik \& Astronomie, Ruhr-Universit\"at Bochum, D-44780 Bochum, Germany}

\noindent
$^{16}${School of Natural Sciences, University of Galway, University Road, Galway, H91 TK33, Ireland}

\noindent
$^{17}${Physics Department, McGill University, Montreal, QC H3A 2T8, Canada}

\noindent
$^{18}${School of Physics, University College Dublin, Belfield, Dublin 4, Ireland}

\noindent
$^{19}${Department of Physics and Astronomy, Barnard College, Columbia University, NY 10027, USA}

\noindent
$^{20}${Institute of Physics and Astronomy, University of Potsdam, 14476 Potsdam-Golm, Germany}

\noindent
$^{21}${Department of Physical Sciences, Munster Technological University, Bishopstown, Cork, T12 P928, Ireland}

\noindent
$^{22}${Department of Physics and Astronomy, University of Alabama, Tuscaloosa, AL 35487, USA}

\noindent
$^{23}${Department of Physics and Astronomy, Purdue University, West Lafayette, IN 47907, USA}

\noindent
$^{24}${Department of Physics, Indiana University Indianapolis, Indianapolis, Indiana 46202, USA}

\noindent
$^{25}${Department of Physics, University of Maryland, Baltimore County, Baltimore MD 21250, USA}

\noindent
$^{26}${Institute for Cosmic Ray Research, University of Tokyo, 5-1-5, Kashiwa-no-ha, Kashiwa, Chiba 277-8582, Japan}
\bibliographystyle{aasjournal.bst}
\bibliography{bib.bib}

\begin{thebibliography}{}
\expandafter\ifx\csname natexlab\endcsname\relax\def\natexlab#1{#1}\fi
\providecommand{\url}[1]{\href{#1}{#1}}
\providecommand{\dodoi}[1]{doi:~\href{http://doi.org/#1}{\nolinkurl{#1}}}
\providecommand{\doeprint}[1]{\href{http://ascl.net/#1}{\nolinkurl{http://ascl.net/#1}}}
\providecommand{\doarXiv}[1]{\href{https://arxiv.org/abs/#1}{\nolinkurl{https://arxiv.org/abs/#1}}}

\bibitem[{{Abdollahi} {et~al.}(2020){Abdollahi}, {Acero}, {Ackermann}, {Ajello}, {Atwood}, {Axelsson}, {Baldini}, {Ballet}, {Barbiellini}, {Bastieri}, {Becerra Gonzalez}, {Bellazzini}, {Berretta}, {Bissaldi}, {Blandford}, {Bloom}, {Bonino}, {Bottacini}, {Brandt}, {Bregeon}, {Bruel}, {Buehler}, {Burnett}, {Buson}, {Cameron}, {Caputo}, {Caraveo}, {Casandjian}, {Castro}, {Cavazzuti}, {Charles}, {Chaty}, {Chen}, {Cheung}, {Chiaro}, {Ciprini}, {Cohen-Tanugi}, {Cominsky}, {Coronado-Bl{\'a}zquez}, {Costantin}, {Cuoco}, {Cutini}, {D'Ammando}, {DeKlotz}, {de la Torre Luque}, {de Palma}, {Desai}, {Digel}, {Di Lalla}, {Di Mauro}, {Di Venere}, {Dom{\'\i}nguez}, {Dumora}, {Fana Dirirsa}, {Fegan}, {Ferrara}, {Franckowiak}, {Fukazawa}, {Funk}, {Fusco}, {Gargano}, {Gasparrini}, {Giglietto}, {Giommi}, {Giordano}, {Giroletti}, {Glanzman}, {Green}, {Grenier}, {Griffin}, {Grondin}, {Grove}, {Guiriec}, {Harding}, {Hayashi}, {Hays}, {Hewitt}, {Horan}, {J{\'o}hannesson}, {Johnson}, {Kamae}, {Kerr}, {Kocevski}, {Kovac'evic'},
  {Kuss}, {Landriu}, {Larsson}, {Latronico}, {Lemoine-Goumard}, {Li}, {Liodakis}, {Longo}, {Loparco}, {Lott}, {Lovellette}, {Lubrano}, {Madejski}, {Maldera}, {Malyshev}, {Manfreda}, {Marchesini}, {Marcotulli}, {Mart{\'\i}-Devesa}, {Martin}, {Massaro}, {Mazziotta}, {McEnery}, {Mereu}, {Meyer}, {Michelson}, {Mirabal}, {Mizuno}, {Monzani}, {Morselli}, {Moskalenko}, {Negro}, {Nuss}, {Ojha}, {Omodei}, {Orienti}, {Orlando}, {Ormes}, {Palatiello}, {Paliya}, {Paneque}, {Pei}, {Pe{\~n}a-Herazo}, {Perkins}, {Persic}, {Pesce-Rollins}, {Petrosian}, {Petrov}, {Piron}, {Poon}, {Porter}, {Principe}, {Rain{\`o}}, {Rando}, {Razzano}, {Razzaque}, {Reimer}, {Reimer}, {Remy}, {Reposeur}, {Romani}, {Saz Parkinson}, {Schinzel}, {Serini}, {Sgr{\`o}}, {Siskind}, {Smith}, {Spandre}, {Spinelli}, {Strong}, {Suson}, {Tajima}, {Takahashi}, {Tak}, {Thayer}, {Thompson}, {Tibaldo}, {Torres}, {Torresi}, {Valverde}, {Van Klaveren}, {van Zyl}, {Wood}, {Yassine}, \& {Zaharijas}}]{4fgl}
{Abdollahi}, S., {Acero}, F., {Ackermann}, M., {et~al.} 2020, The Astrophysical Journal Supplement Series, 247, 33, \dodoi{10.3847/1538-4365/ab6bcb}

\bibitem[{{Abe} {et~al.}(2024){Abe}, {Abe}, {Abhishek}, {Acero}, {Aguasca-Cabot}, {Agudo}, {Alvarez Crespo}, {Antonelli}, {Aramo}, {Arbet-Engels}, {Arcaro}, {Artero}, {Asano}, {Aubert}, {Baktash}, {Bamba}, {Baquero Larriva}, {Baroncelli}, {Barres de Almeida}, {Barrio}, {Batkovic}, {Baxter}, {Becerra Gonz{\'a}ilez}, {Bernardini}, {Bernete Medrano}, {Berti}, {Bhattacharjee}, {Bigongiari}, {Bissaldi}, {Blanch}, {Bonnoli}, {Bordas}, {Brunelli}, {Bulgarelli}, {Burelli}, {Burmistrov}, {Buscemi}, {Cardillo}, {Caroff}, {Carosi}, {Carrasco}, {Cassol}, {Castrej{\'o}n}, {Cauz}, {Cerasole}, {Ceribella}, {Chai}, {Cheng}, {Chiavassa}, {Chikawa}, {Chon}, {Chytka}, {Cicciari}, {Cifuentes}, {Contreras}, {Cortina}, {Costantini}, {Da Vela}, {Dalchenko}, {Dazzi}, {De Angelis}, {de Bony de Lavergne}, {De Lotto}, {de Menezes}, {Del Peral}, {Delgado}, {Delgado Mengual}, {della Volpe}, {Dellaiera}, {Di Piano}, {Di Pierro}, {Di Tria}, {Di Venere}, {D{\'\i}az}, {Dominik}, {Dominis Prester}, {Donini}, {Dorner}, {Doro}, {Eisenberger},
  {Els{\"a}sser}, {Emery}, {Escudero}, {Fallah Ramazani}, {Ferrarotto}, {Fiasson}, {Foffano}, {Freixas Coromina}, {Fr{\"o}se}, {Fukazawa}, {Garcia L{\'o}pez}, {Gasbarra}, {Gasparrini}, {Gavira}, {Geyer}, {Giesbrecht Paiva}, {Giglietto}, {Giordano}, {Gliwny}, {Godinovic}, {Grau}, {Green}, {Green}, {Gunji}, {G{\"u}nther}, {Hackfeld}, {Hadasch}, {Hahn}, {Hassan}, {Hayashi}, {Heckmann}, {Heller}, {Herrera Llorente}, {Hirotani}, {Hoffmann}, {Horns}, {Houles}, {Hrabovsky}, {Hrupec}, {Hui}, {Iarlori}, {Imazawa}, {Inada}, {Inome}, {Ioka}, {Iori}, {Jimenez Martinez}, {Jim{\'e}nez Quiles}, {Jurysek}, {Kagaya}, {Karas}, {Katagiri}, {Kataoka}, {Kerszberg}, {Kobayashi}, {Kohri}, {Kong}, {Kubo}, {Kushida}, {Lainez}, {Lamanna}, {Lamastra}, {Lemoigne}, {Linhoff}, {Longo}, {L{\'o}pez-Coto}, {L{\'o}pez-Moya}, {L{\'o}pez-Oramas}, {Loporchio}, {Lorini}, {Lozano Bahilo}, {Luque-Escamilla}, {Majumdar}, {Makariev}, {Mallamaci}, {Mandat}, {Manganaro}, {Manic{\`o}}, {Mannheim}, {Marchesi}, {Mariotti}, {Marquez}, {Marsella},
  {Mart{\'\i}}, {Martinez}, {Mart{\'\i}nez}, {Mart{\'\i}nez}, {Mas-Aguilar}, {Maurin}, {Mazin}, {Mestre Guillen}, {Micanovic}, {Miceli}, {Miener}, {Miranda}, {Mirzoyan}, {Mizuno}, {Molero Gonzalez}, {Molina}, {Montaruli}, {Moralejo}, {Morcuende}, {Morselli}, {Moya}, {Muraishi}, {Nagataki}, {Nakamori}, {Neronov}, {Nickel}, {Nievas Rosillo}, {Nikolic}, {Nishijima}, {Noda}, {Nosek}, {Novotny}, {Nozaki}, {Ohishi}, {Ohtani}, {Oka}, \& {Okumura}}]{lstcrab}
{Abe}, K., {Abe}, S., {Abhishek}, A., {et~al.} 2024, Astronomy \& Astrophysics, 690, A167, \dodoi{10.1051/0004-6361/202450059}

\bibitem[{Acero {et~al.}(2025)Acero, Aguasca-Cabot, Bernete, Biederbeck, Djuvsland, Donath, Feijen, Fröse, Galelli, Khélifi, Konrad, Kornecki, Linhoff, McKee, Mender, Mohrmann, Morcuende, Olivera-Nieto, Peresano, Pintore, Punch, Regeard, Remy, Roellinghoff, Sinha, Sipőcz, Stapel, Streil, Terrier, Unbehaun, Wong, \& Yu}]{acero_2025_14760974}
Acero, F., Aguasca-Cabot, A., Bernete, J., {et~al.} 2025, Gammapy: Python toolbox for gamma-ray astronomy, v1.3,  Zenodo, \dodoi{10.5281/zenodo.14760974}

\bibitem[{{Ansoldi} {et~al.}(2016){Ansoldi}, {Antonelli}, {Antoranz}, {Babic}, {Bangale}, {Barres de Almeida}, {Barrio}, {Becerra Gonz{\'a}lez}, {Bednarek}, {Bernardini}, {Biasuzzi}, {Biland}, {Blanch}, {Bonnefoy}, {Bonnoli}, {Borracci}, {Bretz}, {Carmona}, {Carosi}, {Colin}, {Colombo}, {Contreras}, {Cortina}, {Covino}, {Da Vela}, {Dazzi}, {De Angelis}, {De Caneva}, {De Lotto}, {de O{\~n}a Wilhelmi}, {Delgado Mendez}, {Di Pierro}, {Dominis Prester}, {Dorner}, {Doro}, {Einecke}, {Eisenacher Glawion}, {Elsaesser}, {Fern{\'a}ndez-Barral}, {Fidalgo}, {Fonseca}, {Font}, {Frantzen}, {Fruck}, {Galindo}, {Garc{\'\i}a L{\'o}pez}, {Garczarczyk}, {Garrido Terrats}, {Gaug}, {Godinovi{\'c}}, {Gonz{\'a}lez Mu{\~n}oz}, {Gozzini}, {Hanabata}, {Hayashida}, {Herrera}, {Hirotani}, {Hose}, {Hrupec}, {Hughes}, {Idec}, {Kellermann}, {Knoetig}, {Kodani}, {Konno}, {Krause}, {Kubo}, {Kushida}, {La Barbera}, {Lelas}, {Lewandowska}, {Lindfors}, {Lombardi}, {Longo}, {L{\'o}pez}, {L{\'o}pez-Coto}, {L{\'o}pez-Oramas}, {Lorenz},
  {Makariev}, {Mallot}, {Maneva}, {Mannheim}, {Maraschi}, {Marcote}, {Mariotti}, {Mart{\'\i}nez}, {Mazin}, {Menzel}, {Miranda}, {Mirzoyan}, {Moralejo}, {Munar-Adrover}, {Nakajima}, {Neustroev}, {Niedzwiecki}, {Nevas Rosillo}, {Nilsson}, {Nishijima}, {Noda}, {Orito}, {Overkemping}, {Paiano}, {Palatiello}, {Paneque}, {Paoletti}, {Paredes}, {Paredes-Fortuny}, {Persic}, {Poutanen}, {Prada Moroni}, {Prandini}, {Puljak}, {Reinthal}, {Rhode}, {Rib{\'o}}, {Rico}, {Rodriguez Garcia}, {Saito}, {Saito}, {Satalecka}, {Scalzotto}, {Scapin}, {Schultz}, {Schweizer}, {Shore}, {Sillanp{\"a}{\"a}}, {Sitarek}, {Snidaric}, {Sobczynska}, {Stamerra}, {Steinbring}, {Strzys}, {Takalo}, {Takami}, {Tavecchio}, {Temnikov}, {Terzi{\'c}}, {Tescaro}, {Teshima}, {Thaele}, {Torres}, {Toyama}, {Treves}, {Ward}, {Will}, \& {Zanin}}]{magiccrab}
{Ansoldi}, S., {Antonelli}, L.~A., {Antoranz}, P., {et~al.} 2016, Astronomy \& Astrophysics, 585, A133, \dodoi{10.1051/0004-6361/201526853}

\bibitem[{Archer {et~al.}(2019)Archer, Benbow, Bird, Brose, Buchovecky, Buckley, Chromey, Cui, Falcone, Feng, Finley, Fortson, Furniss, Gent, Gueta, Hanna, Hassan, Hervet, Holder, Hughes, Humensky, Johnson, Kaaret, Kar, Kelley-Hoskins, Kertzman, Kieda, Krennrich, Kumar, Lang, Lin, McCann, Moriarty, Mukherjee, O’Brien, Ong, Otte, Pandel, Park, Petrashyk, Pohl, Pueschel, Quinn, Ragan, Richards, Roache, Sadeh, Santander, Scott, Sembroski, Shahinyan, Sushch, Tyler, Wakely, Weinstein, Wells, Wilcox, Wilhelm, Williams, Williamson, \& Zitzer}]{archivalpsr}
Archer, A., Benbow, W., Bird, R., {et~al.} 2019, The Astrophysical Journal, 876, 95, \dodoi{10.3847/1538-4357/ab14f4}

\bibitem[{Basu {et~al.}(2021)Basu, Shaw, Antonopoulou, Keith, Lyne, Mickaliger, Stappers, Weltevrede, \& Jordan}]{glitch}
Basu, A., Shaw, B., Antonopoulou, D., {et~al.} 2021, Monthly Notices of the Royal Astronomical Society, 510, 4049, \dodoi{10.1093/mnras/stab3336}

\bibitem[{Bogovalov(2014)}]{bogovalov}
Bogovalov, S.~V. 2014, Monthly Notices of the Royal Astronomical Society, 443, 2197, \dodoi{10.1093/mnras/stu1283}

\bibitem[{{Cogan}(2008)}]{vegref}
{Cogan}, P. 2008, in International Cosmic Ray Conference, Vol.~3, International Cosmic Ray Conference, 1385--1388, \dodoi{10.48550/arXiv.0709.4233}

\bibitem[{{Daugherty} \& {Harding}(1996)}]{polarcap}
{Daugherty}, J.~K., \& {Harding}, A.~K. 1996, Astronomy \& Astrophysics, 120, 107

\bibitem[{{Donath} {et~al.}(2023){Donath}, {Terrier}, {Remy}, {Sinha}, {Nigro}, {Pintore}, {Kh\'elifi}, {Olivera-Nieto}, {Ruiz}, {Br\"ugge}, {Linhoff}, {Contreras}, {Acero}, {Aguasca-Cabot}, {Berge}, {Bhattacharjee}, {Buchner}, {Boisson}, {Carreto Fidalgo}, {Chen}, {de Bony de Lavergne}, {de Miranda Cardoso}, {Deil}, {F\"u\ss{}ling}, {Funk}, {Giunti}, {Hinton}, {Jouvin}, {King}, {Lefaucheur}, {Lemoine-Goumard}, {Lenain}, {L\'opez-Coto}, {Mohrmann}, {Morcuende}, {Panny}, {Regeard}, {Saha}, {Siejkowski}, {Siemiginowska}, {Sip"ocz}, {Unbehaun}, {van Eldik}, {Vuillaume}, \& {Zanin}}]{gammapy:2023}
{Donath}, A., {Terrier}, R., {Remy}, Q., {et~al.} 2023, A\&A, 678, A157, \dodoi{10.1051/0004-6361/202346488}

\bibitem[{{H.~E.~S.~S. Collaboration} {et~al.}(2023){H.~E.~S.~S. Collaboration}, {Aharonian}, {Ait Benkhali}, {Aschersleben}, {Ashkar}, {Backes}, {Barbosa Martins}, {Batzofin}, {Becherini}, {Berge}, {Bernl{\"o}hr}, {Bi}, {B{\"o}ttcher}, {Boisson}, {Bolmont}, {de Bony de Lavergne}, {Borowska}, {Bradascio}, {Breuhaus}, {Brose}, {Brun}, {Bruno}, {Bulik}, {Burger-Scheidlin}, {Bylund}, {Cangemi}, {Caroff}, {Casanova}, {Celic}, {Cerruti}, {Chand}, {Chandra}, {Chen}, {Chibueze}, {Cotter}, {Damascene Mbarubucyeye}, {Djannati-Ata{\"\i}}, {Dmytriiev}, {Egberts}, {Ernenwein}, {Feijen}, {Fiasson}, {Fichet de Clairfontaine}, {Fontaine}, {F{\"u}{\ss}ling}, {Funk}, {Gabici}, {Gallant}, {Ghafourizadeh}, {Giavitto}, {Giunti}, {Glawion}, {Glicenstein}, {Goswami}, {Grolleron}, {Grondin}, {Haerer}, {Haupt}, {Hinton}, {Hofmann}, {Holch}, {Holler}, {Horns}, {Huang}, {Jamrozy}, {Jankowsky}, {Joshi}, {Jung-Richardt}, {Kasai}, {Katarzy{\'n}ski}, {Kh{\'e}lifi}, {Klepser}, {Klu{\v{z}}niak}, {Komin}, {Kosack}, {Kostunin}, {Lang}, {Le
  Stum}, {Lemi{\`e}re}, {Lemoine-Goumard}, {Lenain}, {Leuschner}, {Lohse}, {Luashvili}, {Lypova}, {Mackey}, {Malyshev}, {Malyshev}, {Marandon}, {Marchegiani}, {Marcowith}, {Marinos}, {Mart{\'\i}-Devesa}, {Marx}, {Maurin}, {Meyer}, {Mitchell}, {Moderski}, {Mohrmann}, {Montanari}, {Moulin}, {Muller}, {Murach}, {Nakashima}, {de Naurois}, {Niemiec}, {Noel}, {O'Brien}, {Ohm}, {Olivera-Nieto}, {de Ona Wilhelmi}, {Ostrowski}, {Panny}, {Panter}, {Parsons}, {Peron}, {Pita}, {Prokhorov}, {Prokoph}, {P{\"u}hlhofer}, {Punch}, {Quirrenbach}, {Reichherzer}, {Reimer}, {Reimer}, {Renaud}, {Rieger}, {Rowell}, {Rudak}, {Ruiz-Velasco}, {Sahakian}, {Sailer}, {Salzmann}, {Sanchez}, {Santangelo}, {Sasaki}, {Sch{\"u}ssler}, {Schwanke}, {Shapopi}, {Sinha}, {Sol}, {Specovius}, {Spencer}, {Spir-Jacob}, {Stawarz}, {Steenkamp}, {Steinmassl}, {Steppa}, {Sushch}, {Suzuki}, {Takahashi}, {Tanaka}, {Tavernier}, {Terrier}, {Thorpe-Morgan}, {Tluczykont}, {Tsirou}, {Tsuji}, {van Eldik}, {Vecchi}, {Veh}, {Venter}, {Vink}, {Wagner}, {Werner},
  {White}, {Wierzcholska}, {Wun Wong}, {Yassin}, {Zacharias}, {Zargaryan}, {Zdziarski}, {Zech}, {Zhu}, {Zouari}, {{\.Z}ywucka}, {Zanin}, {Kerr}, {Johnston}, {Shannon}, \& {Smith}}]{hessvela}
{H.~E.~S.~S. Collaboration}, {Aharonian}, F., {Ait Benkhali}, F., {et~al.} 2023, Nature Astronomy, 7, 1341, \dodoi{10.1038/s41550-023-02052-3}

\bibitem[{{Halpern} {et~al.}(2001){Halpern}, {Camilo}, {Gotthelf}, {Helfand}, {Kramer}, {Lyne}, {Leighly}, \& {Eracleous}}]{2001ApJ...552L.125H}
{Halpern}, J.~P., {Camilo}, F., {Gotthelf}, E.~V., {et~al.} 2001, Astrophysical Journal Letters, 552, L125, \dodoi{10.1086/320347}

\bibitem[{{Hobbs} {et~al.}(2006){Hobbs}, {Edwards}, \& {Manchester}}]{2006MNRAS.369..655H}
{Hobbs}, G.~B., {Edwards}, R.~T., \& {Manchester}, R.~N. 2006, Monthly Notices of the Royal Astronomical Society, 369, 655, \dodoi{10.1111/j.1365-2966.2006.10302.x}

\bibitem[{Kuiper \& Hermsen(2015)}]{kuipermwl}
Kuiper, L., \& Hermsen, W. 2015, Monthly Notices of the Royal Astronomical Society, 449, 3827, \dodoi{10.1093/mnras/stv426}

\bibitem[{{Li} \& {Ma}(1983)}]{lima}
{Li}, T.~P., \& {Ma}, Y.~Q. 1983, Astrophysical Journal, 272, 317, \dodoi{10.1086/161295}

\bibitem[{{Lyne} {et~al.}(1993){Lyne}, {Pritchard}, \& {Graham Smith}}]{1993MNRAS.265.1003L}
{Lyne}, A.~G., {Pritchard}, R.~S., \& {Graham Smith}, F. 1993, Monthly Notices of the Royal Astronomical Society, 265, 1003, \dodoi{10.1093/mnras/265.4.1003}

\bibitem[{{Maier} \& {Holder}(2017)}]{edref}
{Maier}, G., \& {Holder}, J. 2017, in International Cosmic Ray Conference, Vol. 301, 35th International Cosmic Ray Conference (ICRC2017), 747, \dodoi{10.22323/1.301.0747}

\bibitem[{Maier {et~al.}(2024)Maier, Holder, McCann, Behera, Duke, Giuri, Skole, Tak, Aliu, Pueschel, Pizlo, Decerpri, Finneagan, Foote, Hughes, Fleischhack, Krawczynski, Prokoph, Grube, Tyler, Berger, Pfrang, Gerard, Beilicke, Kherlakian, Krause, McCutcheon, Nievas-Rosillo, Schroedter, Shayduk, Hakasson, Kelley-Hoskins, Gueta, Ivo, Guenette, Patel, Prado, Welsing, Griffin, Griffiths, O'Brian, Vincent, Vorobiov, Hassan, \& Khassen}]{maier_2024_14283615}
Maier, G., Holder, J., McCann, A., {et~al.} 2024, Eventdisplay: an Analysis and Reconstruction Package for VERITAS, v491.0,  Zenodo, \dodoi{10.5281/zenodo.14283615}

\bibitem[{{Manchester} {et~al.}(2005){Manchester}, {Hobbs}, {Teoh}, \& {Hobbs}}]{atnf}
{Manchester}, R.~N., {Hobbs}, G.~B., {Teoh}, A., \& {Hobbs}, M. 2005, Astrophysical Journal, 129, 1993, \dodoi{10.1086/428488}

\bibitem[{{Nguyen} \& {VERITAS Collaboration}(2015)}]{nguyen_crab}
{Nguyen}, T., \& {VERITAS Collaboration}. 2015, in International Cosmic Ray Conference, Vol.~34, 34th International Cosmic Ray Conference (ICRC2015), 828, \dodoi{10.22323/1.236.0828}

\bibitem[{{Rolke} {et~al.}(2005){Rolke}, {L{\'o}pez}, \& {Conrad}}]{rolke}
{Rolke}, W.~A., {L{\'o}pez}, A.~M., \& {Conrad}, J. 2005, Nuclear Instruments and Methods in Physics Research A, 551, 493, \dodoi{10.1016/j.nima.2005.05.068}

\bibitem[{{Smith} {et~al.}(2023){Smith}, {Abdollahi}, {Ajello}, {Bailes}, {Baldini}, {Ballet}, {Baring}, {Bassa}, {Gonzalez}, {Bellazzini}, {Berretta}, {Bhattacharyya}, {Bissaldi}, {Bonino}, {Bottacini}, {Bregeon}, {Bruel}, {Burgay}, {Burnett}, {Cameron}, {Camilo}, {Caputo}, {Caraveo}, {Cavazzuti}, {Chiaro}, {Ciprini}, {Clark}, {Cognard}, {Corongiu}, {Orestano}, {Crnogorcevic}, {Cuoco}, {Cutini}, {D'Ammando}, {de Angelis}, {DeCesar}, {De Gaetano}, {de Menezes}, {Deneva}, {de Palma}, {Di Lalla}, {Dirirsa}, {Di Venere}, {Dom{\'\i}nguez}, {Dumora}, {Fegan}, {Ferrara}, {Fiori}, {Fleischhack}, {Flynn}, {Franckowiak}, {Freire}, {Fukazawa}, {Fusco}, {Galanti}, {Gammaldi}, {Gargano}, {Gasparrini}, {Giacchino}, {Giglietto}, {Giordano}, {Giroletti}, {Green}, {Grenier}, {Guillemot}, {Guiriec}, {Gustafsson}, {Harding}, {Hays}, {Hewitt}, {Horan}, {Hou}, {Jankowski}, {Johnson}, {Johnson}, {Johnston}, {Kataoka}, {Keith}, {Kerr}, {Kramer}, {Kuss}, {Latronico}, {Lee}, {Li}, {Li}, {Limyansky}, {Longo}, {Loparco}, {Lorusso},
  {Lovellette}, {Lower}, {Lubrano}, {Lyne}, {Maan}, {Maldera}, {Manchester}, {Manfreda}, {Marelli}, {Mart{\'\i}-Devesa}, {Mazziotta}, {McEnery}, {Mereu}, {Michelson}, {Mickaliger}, {Mitthumsiri}, {Mizuno}, {Moiseev}, {Monzani}, {Morselli}, {Negro}, {Nemmen}, {Nieder}, {Nuss}, {Omodei}, {Orienti}, {Orlando}, {Ormes}, {Palatiello}, {Paneque}, {Panzarini}, {Parthasarathy}, {Persic}, {Pesce-Rollins}, {Pillera}, {Poon}, {Porter}, {Possenti}, {Principe}, {Rain{\`o}}, {Rando}, {Ransom}, {Ray}, {Razzano}, {Razzaque}, {Reimer}, {Reimer}, {Renault-Tinacci}, {Romani}, {S{\'a}nchez-Conde}, {Parkinson}, {Scotton}, {Serini}, {Sgr{\`o}}, {Shannon}, {Sharma}, {Shen}, {Siskind}, {Spandre}, {Spinelli}, {Stappers}, {Stephens}, {Suson}, {Tabassum}, {Tajima}, {Tak}, {Theureau}, {Thompson}, {Tibolla}, {Torres}, {Valverde}, {Venter}, {Wadiasingh}, {Wang}, {Wang}, {Wang}, {Weltevrede}, {Wood}, {Yan}, {Zaharijas}, {Zhang}, \& {Zhu}}]{3pc}
{Smith}, D.~A., {Abdollahi}, S., {Ajello}, M., {et~al.} 2023, Astrophysical Journal, 958, 191, \dodoi{10.3847/1538-4357/acee67}

\bibitem[{van Scherpenberg {et~al.}(2019)van Scherpenberg, Mirzoyan, Vovk, Peresano, Zarić, Temnikov, Godinović, \& Besenrieder}]{vanscherpenberg2019searchingvariabilitycrabnebula}
van Scherpenberg, J., Mirzoyan, R., Vovk, I., {et~al.} 2019, Searching for Variability of the Crab Nebula Flux at TeV Energies using MAGIC Very Large Zenith Angle Observations.
\newblock \doarXiv{1909.04356}

\bibitem[{{VERITAS Collaboration} {et~al.}(2011){VERITAS Collaboration}, {Aliu}, {Arlen}, {Aune}, {Beilicke}, {Benbow}, {Bouvier}, {Bradbury}, {Buckley}, {Bugaev}, {Byrum}, {Cannon}, {Cesarini}, {Christiansen}, {Ciupik}, {Collins-Hughes}, {Connolly}, {Cui}, {Dickherber}, {Duke}, {Errando}, {Falcone}, {Finley}, {Finnegan}, {Fortson}, {Furniss}, {Galante}, {Gall}, {Gibbs}, {Gillanders}, {Godambe}, {Griffin}, {Grube}, {Guenette}, {Gyuk}, {Hanna}, {Holder}, {Huan}, {Hughes}, {Hui}, {Humensky}, {Imran}, {Kaaret}, {Karlsson}, {Kertzman}, {Kieda}, {Krawczynski}, {Krennrich}, {Lang}, {Lyutikov}, {Madhavan}, {Maier}, {Majumdar}, {McArthur}, {McCann}, {McCutcheon}, {Moriarty}, {Mukherjee}, {Nu{\~n}ez}, {Ong}, {Orr}, {Otte}, {Park}, {Perkins}, {Pizlo}, {Pohl}, {Prokoph}, {Quinn}, {Ragan}, {Reyes}, {Reynolds}, {Roache}, {Rose}, {Ruppel}, {Saxon}, {Schroedter}, {Sembroski}, {{\c{S}}ent{\"u}rk}, {Smith}, {Staszak}, {Te{\v{s}}i{\'c}}, {Theiling}, {Thibadeau}, {Tsurusaki}, {Tyler}, {Varlotta}, {Vassiliev}, {Vincent},
  {Vivier}, {Wakely}, {Ward}, {Weekes}, {Weinstein}, {Weisgarber}, {Williams}, \& {Zitzer}}]{vtscrab}
{VERITAS Collaboration}, {Aliu}, E., {Arlen}, T., {et~al.} 2011, Science, 334, 69, \dodoi{10.1126/science.1208192}

\end{thebibliography}
\end{document}